# 复杂噪音条件下基于抗差容积卡尔曼滤波的发电机动态状态估计

李扬[1]　李京[1]　陈亮[2]　李国庆[1]

（1 东北电力大学电气工程学院，吉林省　吉林市　132012;
2 国网河北省电力有限公司经济技术研究院，河北省 石家庄市 050022;）

**摘要**　容积卡尔曼滤波(cubature Kalman filter, CKF)在非线性动态状态估计领域有着良好的估计效果。但由于容积卡尔曼滤波缺乏对量测噪音特性的在线自适应能力，其对不良数据和非高斯白噪音的处理效果并不理想。为了解决当量测量统计特性偏离先验统计特性时，容积卡尔曼滤波算法性能下降和发散的问题，本文通过将抗差估计理论中的M-估计理论与容积卡尔曼滤波相结合，提出抗差容积卡尔曼滤波 (robust CKF，RCKF)算法，并将其尝试应用于复杂噪声条件下的发电机动态状态估计中。IEEE-9节点系统和新英格兰16机68节点系统的仿真结果表明：在不同量测噪音且量测量存在不良数据的复杂噪音条件下，与传统CKF算法相比，所提RCKF算法均有有更好的估计精度和收敛能力，并能有效消除不良数据对估计效果的影响。

# Dynamic State Estimation of Synchronous Machines Using Robust Cubature Kalman Filter Against Complex Measurement Noise Statistics

*Li Yang* [1], *Li Jing* [1], *Chen Liang* [2], *Li Guoqing* [1]

(1 School of Electrical Engineering, Northeast Electric Power University, Jilin 132012, Jilin Province, China;
2 State Grid Hebei Economic Research Institute, Shijiazhuang 050022, Hebei Province, China;）

**Abstract**　Cubature Kalman Filter (CKF) has good performance when handling nonlinear dynamic state estimations. However, it cannot work well in non-Gaussian noise and bad data environment due to the lack of auto-adaptive ability to measure noise statistics on line. In order to address the problem of behavioral decline and divergence when measure noise statistics deviate prior noise statistics, a new robust CKF (RCKF) algorithm is developed by combining the Huber's M-estimation theory with the classical CKF, and thereby it is proposed to coping with the dynamic state estimation of synchronous generators in this study. The simulation results on the IEEE-9 bus system and New England 16-machine-68-bus system demonstrate that the estimation accuracy and convergence of the proposed RCKF are superior to those of the classical CKF under complex measurement noise environments including different measurement noises and bad data, and that the RCKF is capable of effectively eliminating the impact of bad data on the estimation effects.

## 0 引言

精确、可靠的发电机动态状态量对电力系统的实时监测和控制至关重要[1,2]。但是，大部分发电机状态量无法直接测得，而且传统的数据采集与监控系统(supervisory control and data acquisition，SCADA)只能提供稳态、低采样频率的非同步量测量，无法捕获系统的动态信息[3]。随着在系统中广泛配置相量测量单元(phasor measurement units，PMUs)，利用同步的PMU

国家重点研发计划（2017YFB0902401），国家自然科学基金（51077097）。

数据估计发电机的所有状态量已成为可能[4]。但是，由于生产环境、运行状态等原因，量测量中不可避免地出现大量误差和不良数据。因此，研究考虑机电暂态过程中的发电机动态状态估计（dynamic state estimation，DSE）方法，成为了国内外研究热点[1,2,5]。

扩展卡尔曼滤波(extended Kalman filter, EKF)是迄今为止应用最广泛的非线性状态估计方法之一[6,7]，它通过将非线性问题泰勒展开，然后通过一阶线性截断将非线性问题线性化[8]。由于忽略了高次项，在解决非线性强的问题时，EKF 的状态估计精度较低[9]。无迹卡尔曼滤波（unscented Kalman filter, UKF）利用无迹变换避免了将非线性问题线性化，且无需计算雅格比矩阵。但是，数值稳定性和维数灾难等问题一定程度上限制了 UKF 的应用[10-13]。

容积卡尔曼滤波是一种可以应用于高维度系统的非线性动态状态估计方法。CKF 以球面-径向规则为核心通过球面-径向原则获得的容积点，利用容积点计算后验概率密度函数, 解决了存在于 EKF 和 UKF 中的偏差和维度问题。对于服从高斯白噪音分布的非线性系统，CKF 可以提供更加准确的状态估计结果[14-16]。

基于 CKF 的发电机动态状态估计，需要确定发电机的数学模型、先验的量测噪音和过程噪音。在实际运算中, CKF 先在预报步求得状态量在 $k$ 时刻的预报值,之后利用量测量的预报值进行修正，最终获得可靠的估计值。

文献[15]证明了在发电机动态状态估计问题中，CKF 估计精度，收敛能力等方面优于 UKF。然而，在发电机实际运行时，由于生产环境，运行状态等原因，量测量中不可避免地出现大量误差和不良数据，这些不准确的量测量会使量测噪音统计特性偏离先验特性[17,18]。另一方面，实测 PMU 统计数据显示发电机的量测噪音并不一定服从高斯白噪音分布[19]。这些因素会导致应用于 CKF 的发电机动态状态估计结果的精度下降。

为了使得容积卡尔曼滤波对量测噪音特性具有在线自适应能力，提高其在面对不良数据和非高斯白噪音时的估计效果，本文将传统 CKF 和 M-估计理论相结合，提出基于抗差 CKF 的发电机动态状态估计，使其能够探测量测量中的粗差影响，进而实现量测噪音统计特性的在线自适应调节[20]。最后，以 IEEE-9 节点系统和新英格兰 16 机 68 节点系统为例，验证所提方法在考虑噪音不服从高斯白噪音分布及存在不良数据等复杂噪音条件下的有效性。

## 1 发电机动态状态估计模型

### 1.1 非线性动态系统的数学模型

针对发电机动态状态估计问题，需要确定的非线性数学模型的状态方程和量测方程的形式如下：

$$\begin{cases} x_{k+1} = F(x_k, u_k, v_k) \\ z_{k+1} = H(x_{k+1}, u_{k+1}, w_{k+1}) \end{cases} \tag{1}$$

上式中，$x$ 为状态向量，$u$ 为控制向量，z 为量测向量，$v$ 和 $w$ 分别为系统噪音向量和量测噪音向量，其取值服从正态分布且均值为 0；下标 $k$ 为时刻。

### 1.2 发电机的数学模型

在发电机动态状态估计领域，一般选用发电机四阶数学模型作为状态量模型进行仿真计算。

$$\begin{cases} \dot{\delta} = \omega - 1 \\ \dot{\omega} = \dfrac{1}{T_J}\left[T_m - T_e - D(\omega - 1)\right] \\ \dot{E}_q' = \dfrac{1}{T_{d0}'}\left[E_f - E_q' - \left(X_d - X_d'\right)I_d\right] \\ \dot{E}_d' = \dfrac{1}{T_{q0}'}\left[-E_d' + \left(X_q - X_q'\right)I_q\right] \end{cases} \tag{2}$$

$$\begin{cases} I_d = \dfrac{E_q' - U_t\cos(\delta - \varphi)}{X_d'} \\ I_q = \dfrac{U_t\sin(\delta - \varphi) - E_d'}{X_q'} \end{cases} \tag{3}$$

式中，$\delta$ 是发电机功角即励磁电动势和端电压的夹角，$\omega$ 是发电机角速度标幺值，$E_q'$ 是发电机交轴瞬变电动势，$E_d'$ 是发电机直轴瞬变电动势；$T_m$、$T_e$、$D$ 分别是发电机机械转矩、发电机电磁转矩以及阻尼系数；$T'_{d0}$ 是直轴暂态时间常数，$T'_{q0}$ 是交轴暂态时间常数，$E_f$ 为励磁电动势；$X_d$、$X_q$、$X'_d$、$X'_q$ 分别是直轴同步电抗、交轴同步电抗、直轴暂态电抗，交轴暂态电抗。$U_t$ 和 $\varphi$ 为发电机电压幅值和相角；$I_d$、$I_q$ 分别是发电机输出电流的直轴分量和交轴分量。

发电机量测方程如下：

$$\begin{cases} \omega^z = \omega \\ \delta^z = \delta \\ P_e^z = \dfrac{U_t^2}{2}\sin(2\delta - 2\varphi)\left(\dfrac{1}{X_q'} - \dfrac{1}{X_d'}\right) + \\ \qquad \dfrac{U_t^2\sin(\delta - \varphi)E_q'}{X_d'} - \dfrac{U_t^2\cos(\delta - \varphi)E_d'}{X_q'} \end{cases} \tag{4}$$

式中，$\delta^z$ 和 $\omega^z$ 分别为发电机功角和角速度的量测量，$P_e^z$ 是发电机电磁功率量测量。$\delta^z$ 和 $\omega^z$ 可以直接通过 PMU 获得，电磁功率 $P_e^z$ 则可以通过计算获得。

发电机量测噪音方差阵 $R_{k+1}$ 为：

$$R_{k+1} = diag\left(\sigma_{\delta_z}^2, \sigma_{\omega_z}^2, \sigma_{Pe_z}^2\right) \quad (5)$$

发电机功角量测方差 $\sigma_{\delta z}{}^2$=4°，角速度量测方差 $\sigma_{\omega z}{}^2=1\times10^{-6}$，电磁功率量测方差的计算公式为：

$$P_{e_z} = \left(\frac{\partial P_e}{\partial U}\right)^2 \sigma_{U_t}^2 + \left(\frac{\partial P_e}{\partial \varphi}\right)^2 \sigma_\varphi^2 \quad (6)$$

式中，$\sigma_{U_t}$=0.2%，$\sigma_\phi$=0.2°。

在发电机动态状态估计过程中，系统噪音主来源于$\delta_t$和$U_t$的量测误差，该量测误差随系统状态方程进行传递，最终导致量测噪音的产生。$k$时刻的系统噪音方差阵$Q_k$为：

$$Q_k = diag\left(\sigma_{m\delta_k}^2, \delta_{m\xi_k}^2, \delta_{mE_{qk}'}^2, \delta_{mE_{dk}'}^2\right) \quad (7)$$

式中，$\sigma_{m\delta k}{}^2$，$\sigma_{m\omega k}{}^2$，$\sigma^2_{mEq'k}$，$\sigma^2_{mEd'k}$分别是在$k$时刻发电机状态量$\delta, \omega, E_d', E_q'$的系统噪音方差。具体数值可通过误差传递公式计算：

$$\sigma^2 = \left(\frac{\partial F}{\partial U_t}\right)^2 \sigma_{U_t}^2 + \left(\frac{\partial F}{\partial \delta_t}\right)^2 \sigma_{\delta_t}^2 \quad (8)$$

式中，$\sigma_{ut}$和$\sigma_{\phi\tau}$分别为发电机出口电压幅值和相角的量测误差标准差。

综合式(2)至式（8）可得发电机的非线性系统方程中各个函数如下：

$$\begin{cases} x = \left[\delta, \omega, E_d', E_q'\right] \\ z = \left[\delta^z, \omega^z, P_e^z\right] \\ u = \left[T_m, E_f, U_t, \delta\right] \\ v_k = Q_k \\ w_{k+1} = R_{k+1} \end{cases} \quad (9)$$

## 2 基于抗差 CKF 的发电机动态状态估计

### 2.1 容积卡尔曼滤波

CKF 算法分为两部分:

1) 状态预报步和滤波修正步。在状态预报步中，CKF 根据球面-径向产生一组等权值的状态量容积点，之后利用系统方程获得下一时刻状态量的预报值，然后通过计算获得预报误差方差矩阵;

2) 在滤波步中，通过量测值完成对状态量预报值的修正，得到更为准确的估计值。具体过程如下：

状态预报步：

假设抑制 $k$ 时刻的后验概率分布，得到状态向量的估计误差方差阵 $P_{k/k}$。

$$P_{k/k} = S_{k/k} S_{k/k}{}^T \quad (10)$$

计算状态量容积点

$$X_{i,k/k} = S_{k/k}\zeta_i + \hat{x}_{k/k} \quad (11)$$

式中，$i$ 为状态量容积点的编号。

通过状态方程获得状态量预报值容积点

$$X_{i,k/k}^* = F\left(X_{i,k/k}, u_k\right) \quad (12)$$

通过加权求和获得状态量预报值

$$\hat{x}_{k+1/k} = \frac{1}{2n}\sum_{i=1}^{2n} X_{i,k+1/k}^* \quad (13)$$

计算状态量预报误差方差阵

$$P_{k+1/k} = \frac{1}{2n}\sum_{i=1}^{2n} X_{i,k+1/k}^* X_{i,k+1/k}^{*T} - \hat{x}_{k+1/k} \hat{x}_{k+1/k}^T + Q_k \quad (14)$$

滤波修正步：

计算预报误差方差阵的平方根矩阵

$$P_{k+1/k} = S_{k+1/k} S_{k+1/k}^T \quad (15)$$

计算状态量预报值容积点

$$X_{i,k+1/k} = S_{k+1/k}\zeta_i + \hat{x}_{k+1/k} \quad (16)$$

通过量测方程获得量测量预报值容积点

$$Z_{i,k+1/k} = H\left(X_{i,k+1/k}, u_k\right) \quad (17)$$

通过加权求和获得量测量预报值

$$\hat{z}_{k+1/k} = \frac{1}{2n}\sum_{i=1}^{2n} Z_{i,k+1/k} \quad (18)$$

计算量测量误差方差阵

$$P_{zz,k+1/k} = \frac{1}{2n}\sum_{i=1}^{2n} Z_{i,k+1/k} Z_{i,k+1/k}^T - \hat{z}_{k+1/k}\hat{z}_{k+1/k}^T + R_{k+1} \quad (19)$$

计算交叉误差方差阵

$$P_{xz,k+1/k} = \frac{1}{2n}\sum_{i=1}^{2n} X_{i,k+1/k} Z_{i,k+1/k}^T - \hat{x}_{k+1/k}\hat{z}_{k+1/k}^T \quad (20)$$

计算滤波增益

$$W_{k+1} = P_{xz,k+1/k} P_{zz,k+1/k}^{-1} \quad (21)$$

获得状态量估计值

$$\hat{x}_{k+1/k+1} = \hat{x}_{k+1/k} + W_{k+1}\left(z_{k+1} - \hat{z}_{k+1/k}\right) \quad (22)$$

更新下一时刻状态量估计误差方差矩阵

$$P_{k+1/k+1}=P_{k+1/k}-W_{k+1}P_{zz,k+1/k}W_{k+1}^{T} \quad (23)$$

## 2.2 抗差 CKF

拥有精确的系统模型和噪音统计特性是传统的 CKF 获得良好的估计效果的前提。在发电机动态状态估计中，发电机模型是准确的。但在实际生产中，由于环境等因素的影响，PMU 的量测数据中会出现不良数据且噪音分布并不一定符合高斯白噪音分布。这些因素将导致量测误差方差阵 $R_{k+1}$ 与实际的误差不符，从而导致在修正步中 CKF 无法完成对状态量预报值的准确修正，最终使状态估计效果不理想。

通过抗差理论中的稳健 M 估计，探测量测量中的不良数据对状态估计的影响程度，以此对量测噪音统计特性进行实时的更新，使 CKF 的量测噪音统计特性获得自适应的能力，能够及时地对量测噪音统计特性的变化做出反应，获得相应的量测误差方差阵，以此构建了一种具有对量测噪音统计特性在线调整能力的抗差 CKF。该方法可以在发电机量测量出现不良数据和噪音不服从高斯白噪音分布时，获得准确的状态估计结果。

抗差 CKF 的基本过程大体与 CKF 相同，不过需要运用抗差理论获得修正后的量测噪音方差阵并用其替换掉(19)步中的修正前量测噪音方差阵。

### 2.2.1 抗差理论

$\bar{R}_{k+1}$ 为量测噪音方差阵，$\bar{R}_{k+1}$ 为修正后的量测噪音方差阵，抗差 M 估计方法中的等价权矩阵为$\bar{P}$：

$$\bar{R}_{k+1}=\bar{P}^{-1} \quad (24)$$

文采用 Huber 法计算等价权矩阵$\bar{P}$，如下：

$$\bar{p}_{i,i}=\begin{cases}\dfrac{1}{\sigma_{i,i}}, & \left(\left|\dfrac{r_i}{\sigma_{ri}}\right|=\left|r_i^{'}\right|\le c\right)\\[2ex] \dfrac{c}{\sigma_{i,i}\left|r_{i,i}\right|}, & \left(\left|r_i^{'}\right|>c\right)\end{cases} \quad (25)$$

$$\bar{p}_{i,j}=\begin{cases}\dfrac{1}{\sigma_{i,j}}, & \left(\left|r_i^{'}\right|\le c,\left|r_i^{'}\right|\le c\right)\\[2ex] \dfrac{c}{\sigma_{i,j}\max\left(\left|r_i^{'}\right|,\left|r_j^{'}\right|\right)}, & \left(\left|r_i^{'}\right|>c,\left|r_j^{'}\right|>c\right)\end{cases} \quad (26)$$

式中，$\bar{p}_{i,i}$ 和$\bar{p}_{i,j}$ 分别是$\bar{P}$的对角元素和非对角元素；$\sigma_{i,i}$ 和$\sigma_{i,j}$ 为量测噪音方差阵 $R_k$ 的对角元素和非对角元素。由于发电机动态状态估计中的量测噪音方差阵为对角矩阵，非对角元素为零，因此$\sigma_{i,j}$ 取零。$r_i$ 为量测量 $Z_i$ 的对应残差分量，$r_i'$是相应的标准残差分量，$\sigma_i$ 为 $r_i$ 的均方差；$c$ 为一个给定的常数，其取值范围为 1.3 到 2.0 之间，文中其值为 1.5。

$\sigma_{ri}$ 和 $r_i$ 的取值规则如下：

$$\sigma_{ri}=(P_{\mathrm{zz,k+1/k}})_{i,i} \quad (27)$$

$$r_i=(z_{\mathrm{k+1}}-\hat{z}_{\mathrm{k+1/k}})_i \quad (28)$$

$P_{zz,k+1/k}$ 为修正前的量测量误差总方差。

### 2.2.2 估计过程

在电力系统动态过程中，由于其拓扑结构发生突变，系统状态方程相应地出现改变。这种改变使动态状态估计的精确度大幅下降。为消除拓扑结构突变的影响，研究者们利用 PMU 实时获得发电机出口电压向量，以此解列发电机与系统其余部分。同时，发电机的动态方程在系统动态过程中是稳定不变的，这就为发电机的动态状态估计提供了基础。选用发电机的四阶动态方程作为发电机动态状态估计的状态方程，考虑量测噪音和过程噪音的影响，从而完成基于抗差容积卡尔曼滤波的发电机动态状态估计。

根据式(9)，可获得各函数如下：

$x=[\delta, \omega, E'_d, E'_q]$；$u=[T_m, E_f, U_t, \varphi]$；$z=[\delta^z, \omega^z, P_e^z]$；$U_t$ 和$\varphi$为 PMU 量测值。

基于抗差 CKF 完成发电机的动态状态估计的基本过程：

1）预报步

(1) $k$ 时刻发电机状态量估计值和估计误差方差矩阵分别为 $\hat{x}_k=\left[\delta_k,\omega_k,E_{d,k}^{'},E_{q,k}^{'}\right]^T$ 和 $P_{k,k}$。随后根据(10)对 $k$ 时刻的估计误差方差矩阵进行 Cholesky 分解，获得 $P_{k/k}$ 的平方根矩阵 $S_{k/k}$。

(2) 根据式(11)，通过球面-径向原则获得发电机状态量容积点。根据(12)获得发电机状态量容积点在 $k$+1 时刻的预报值 $\hat{x}_{k+1/k}=\left[\delta_{k+1/k},\omega_{k+1/k},E_{d,k+1/k}^{'},E_{q,k+1/k}^{'}\right]^T$。

(3) 根据式(14)获得发电机状态量预报误差方差阵 $P_{k+1,k}$。

2）滤波步

(1) 根据式(15)和式(16)计算预报误差方差的平方根矩阵和状态量预报值容积点。

(2) 根据式(17)和(18)获得量测量预报值 $\hat{z}_{k+1/k}=\left[\delta_{k+1/k}^{z},\omega_{k+1/k}^{z},P_{e,k+1/k}^{z}\right]$。

(3) 利用式(19)计算量测误差方差阵，利用式(24)

至式(26)对量测误差方差阵进行修正。

(4) 根据式(20)至(22),获得 $k+1$ 时刻的发电机状态量估计值：

$$\hat{x}_{k+1/k+1}=\left[\delta_{k+1/k+1},\omega_{k+1/k+1},E'_{d,k+1/k+1},E'_{q,k+1/k+1}\right]^T$$

根据式(19)获得 $k+1$ 时刻的状态量估计误差方差矩阵 $P_{k+1,k+1}$。

## 3 仿真分析

### 3.1 仿真系统设置

在 MATLAB 环境下进行仿真，仿真结果作为真值，在真值的基础上分别叠加高斯白噪音，高斯有偏噪音，拉普拉斯噪音和柯西噪音作为量测量。所有仿真均在配置为 Intel Core i5-4590 3.3 GHz 处理器、8 GB 内存的 PC 机上实现。仿真基准频率为 50Hz，采样步长为 0.02s；PMU 量测误差标准差设置：发电机功角标准差是 2°，角速度标准差是 0.1%，出口电压的相角和幅值标准差是 0.1° 和 0.1%。

为验证抗差容积卡尔曼滤波对不良数据的鲁棒性，在发电机电角速度的 PMU 量测中人为设置不良数据点。考虑不良数据点是单个数据和连续数据两种情况，在 IEEE9 节点系统中，在 $t$=6s 叠加单个不良数据点，在 $t$=12s 开始连续叠加 10 个不良数据点；在新英格兰 68 节点系统中，在 $t$=2s 叠加单个不良数据点，在 $t$=3s 开始连续叠加 10 个不良数据点。

### 3.2 噪音模型和参数设置

**3.2.1** 高斯噪音

$$f(x)=\frac{1}{\sigma\sqrt{2\pi}}e^{-\frac{(l-\mu)^2}{2\sigma^2}} \qquad (29)$$

式中，$\mu$为 $l$ 的平均值，$\sigma$为 $l$ 的标准差。

**3.2.2** 拉普拉斯噪音

$$r_{Laplace}=m-s\,\mathrm{sgn}(U)\ln(1-|U|) \qquad (30)$$

式中，$m$ 为平均值，$s$ 为刻度参数，$U$ 为在服从均匀分布的采样区间内的随机数。

**3.2.3** 柯西噪音

$$r_{Cauchy}=a+b\tan\left(\pi\left(U_2-0.5\right)\right) \qquad (31)$$

式中，$\alpha$为位置参数，$b$ 为刻度参数，$U_2$ 为在服从均匀分布的采样区间内的随机数。

**3.2.4** 参数设置

算例中所用的噪声类型与参数设置如表 1 所示。

**表 1 噪音类型与参数设置**

Tab.1 Noise Type and Parameters

| 噪音类型 | 参数 | 标准差 | 均值 |
|---|---|---|---|
| 高斯白噪音 | 功角(°) | 2 | 0 |
| | 角速度(%) | 0.1 | 0 |
| 高斯有偏噪音 | 功角(°) | 2 | 20 |
| | 角速度(%) | 0.1 | 1 |
| 拉普拉斯噪音 | 功角(°) | 2 | 20 |
| | 角速度(%) | 0.1 | 1 |
| 柯西噪音 | 功角(°) | 2 | 20 |
| | 角速度(%) | 0.1 | 1 |

### 3.3 评估指标

建立发电机 $\varepsilon_1$ 动态状态估计的量化评估指标及滤波系数和估计误差总方差 $\varepsilon_2$，具体计算公式如下[15]：

$$\varepsilon_1=\sqrt{\frac{\sum_{i=1}^{S}\left(\hat{x}_i-x_i^t\right)^2}{\sum_{i=1}^{S}\left(x_i^z-x_i^t\right)^2}} \qquad (32)$$

$$\varepsilon_2=\sqrt{\frac{1}{S}\sum_{i=1}^{S}\left(\frac{\hat{x}_i-x_i^t}{x_i^t}\right)^2} \qquad (33)$$

式中，$\hat{x}_i$，$x_i^t$ 和 $x_i^z$ 分别为估计值，真实值和测量值，$S$ 为采样点数。估计指标 $\varepsilon_1$ 可评价在同一种噪音特性时，不同状态估计方法的滤波性能；$\varepsilon_2$ 可衡量同一状态估计方法在不同噪音特性下的滤波性能。

### 3.4 IEEE 9 节点系统

将所提算法用于 IEEE 9 节点系统进行仿真，仿真设置如下：在 $t$=1.2s 时节点 5 发生三相金属性短路故障，随后故障被断路器切除，仿真持续时间为 20s。

**3.4.1** 仿真结果

为了不失一般性，以发电机 2 为例，针对不同噪声环境的仿真结果展示如下。

**（1）高斯白噪音**

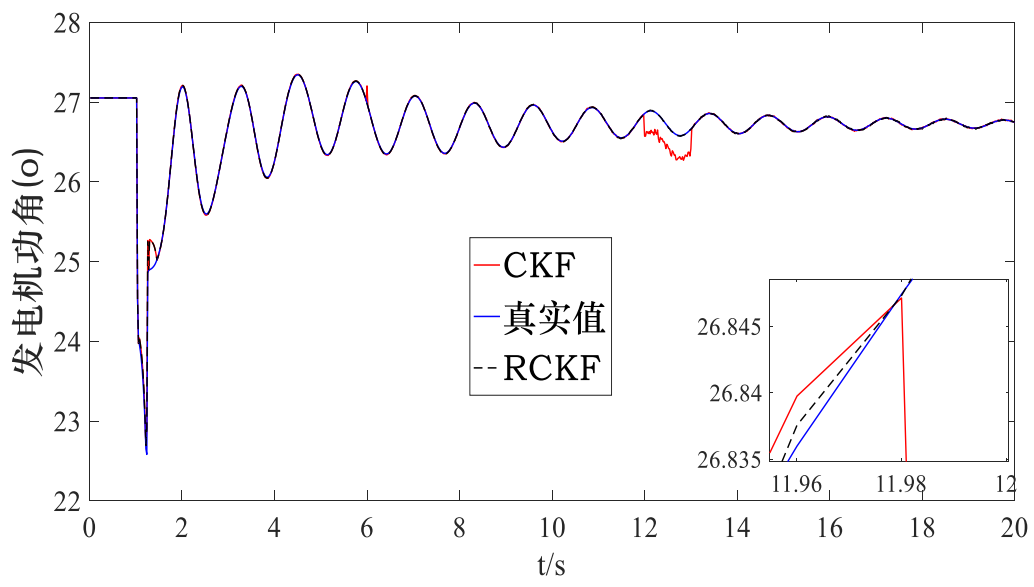

图 1 发电机功角

Fig.1 Estimation of generation power angle

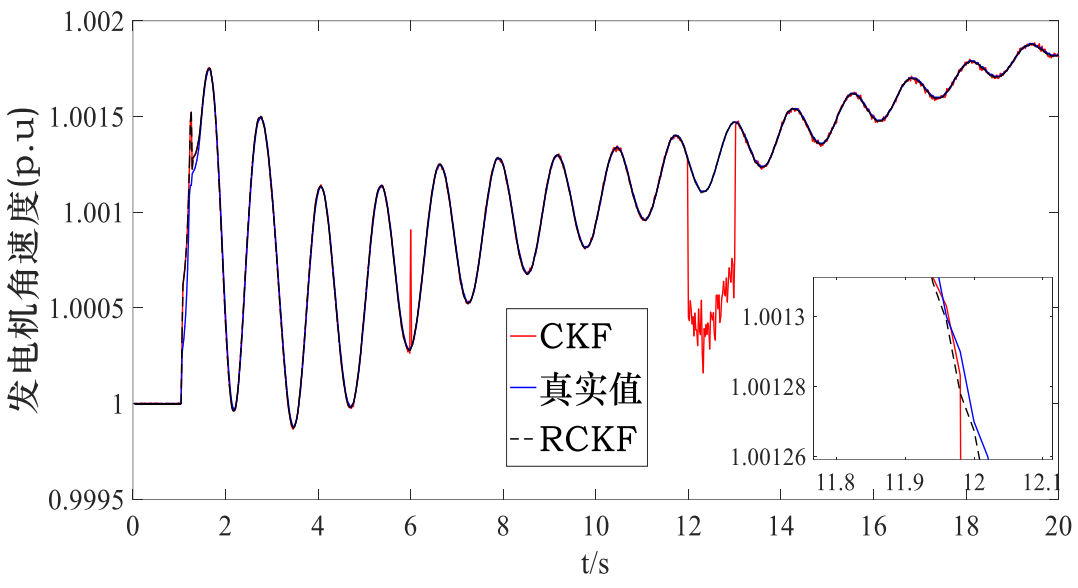

图 2　发电机角速度

Fig.2　Estimation of generation angular velocity

**（2）　高斯有偏噪音**

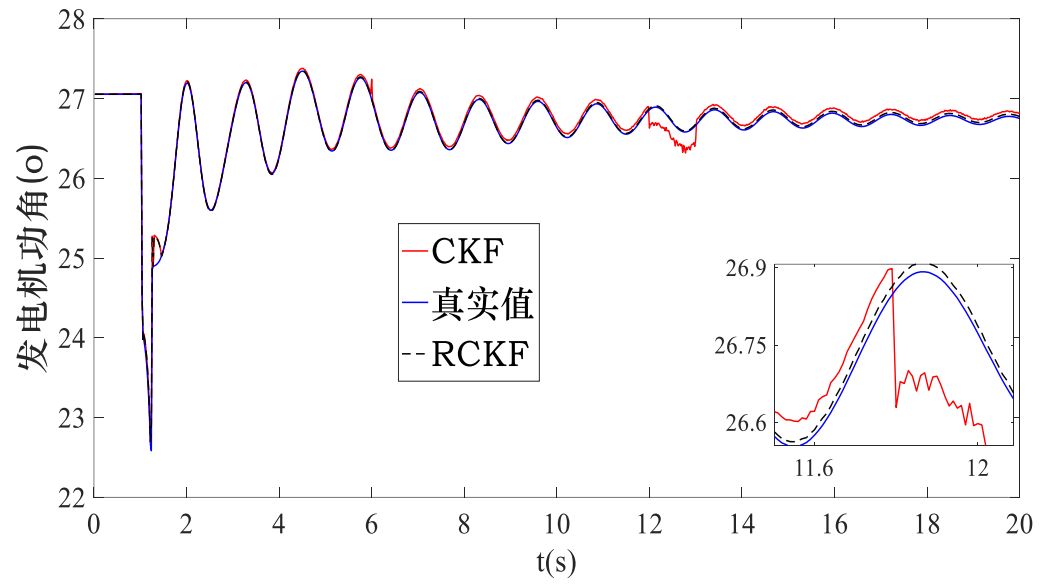

图 3　发电机功角

Fig.3　Estimation of generation power angle

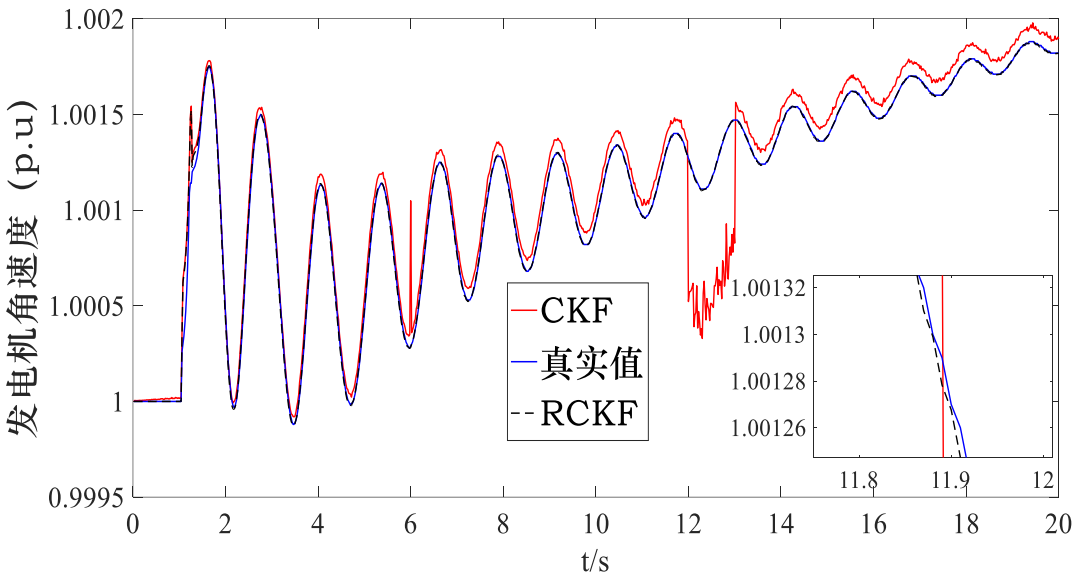

图 4　发电机角速度

Fig.4　Estimation of generation angular velocity

**（3）　拉普拉斯噪音**

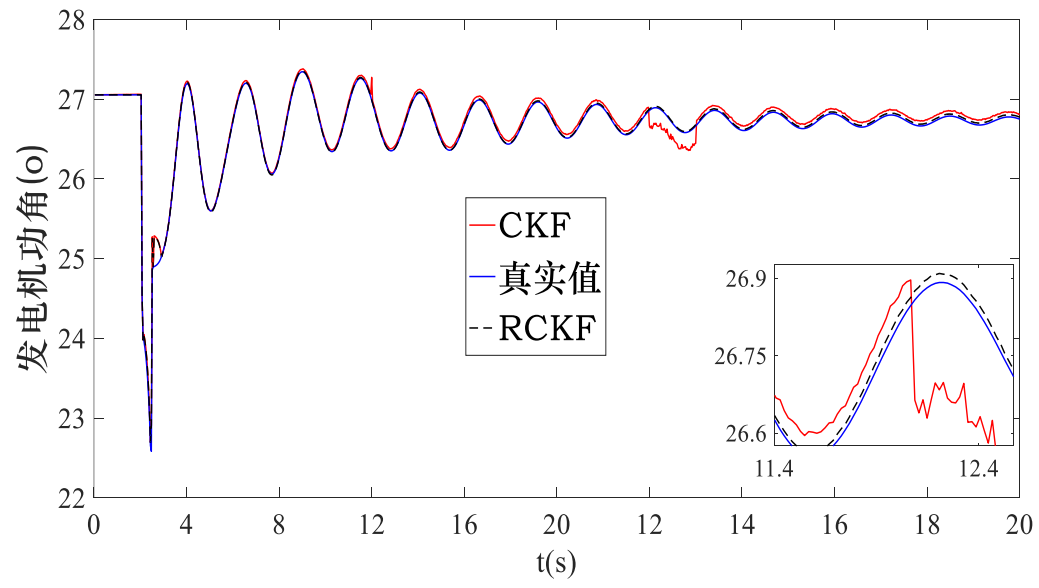

图 5　发电机功角

Fig.5　Estimation of generation power angle

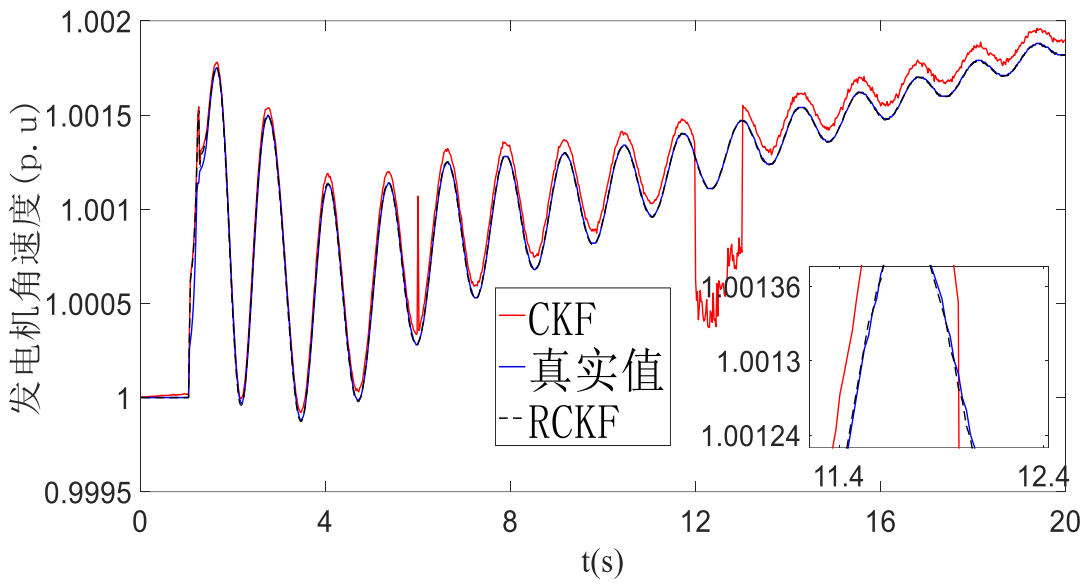

图 6　发电机角速度

Fig.6　Estimation of generation angular velocity

**（4）　柯西噪音**

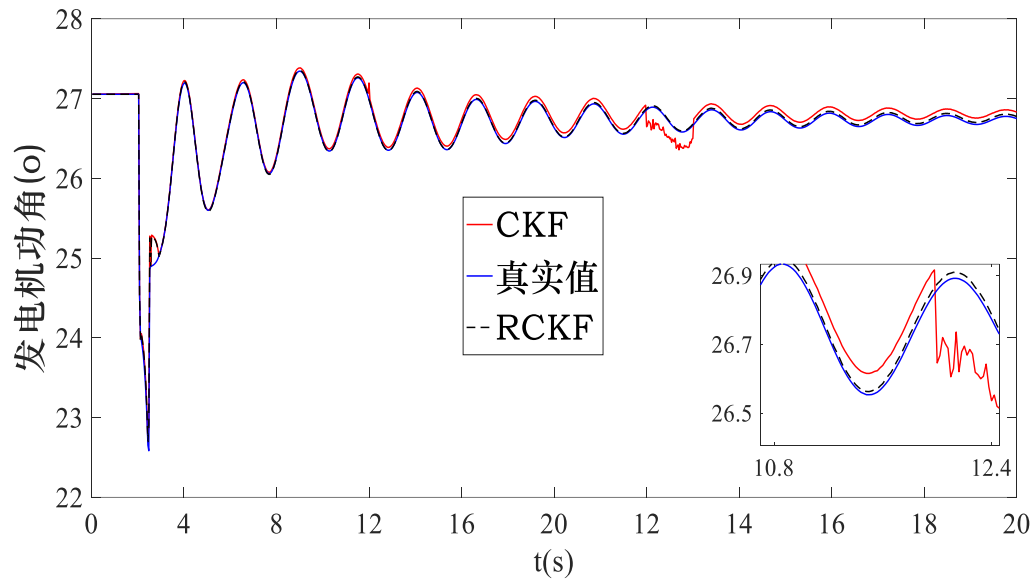

图 7　发电机功角

Fig.7　Estimation of generation power angle

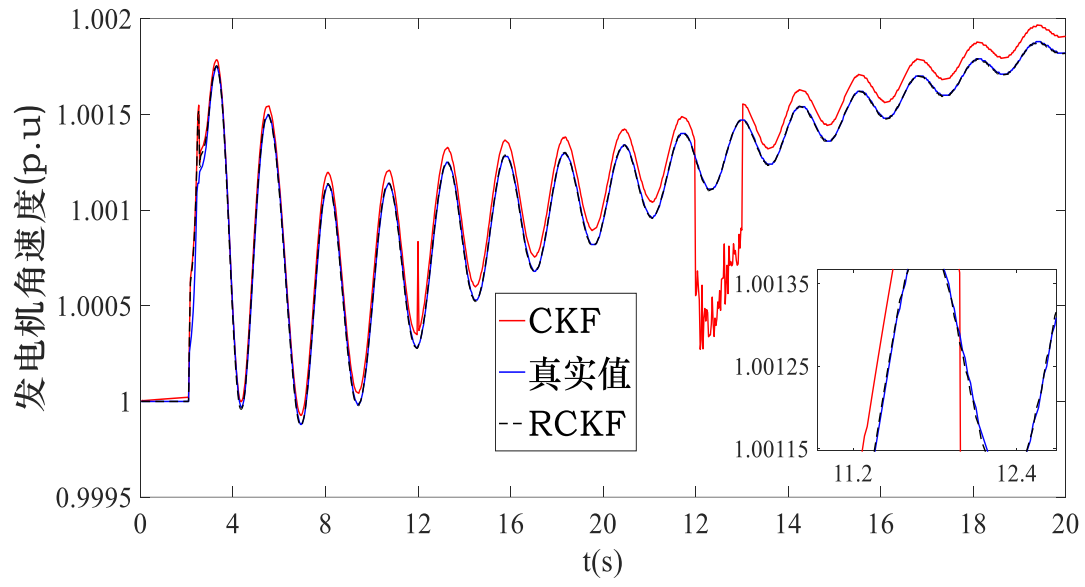

图 8　发电机角速度

Fig.8　Estimation of generation angular velocity

图 1,3,5,7 显示了在不同类型的量测噪音环境下，发电机功角的动态状态估计结果；图 2,4,6,8 显示了在不同类型的量测噪音环境下，发电机角速度的动态状态估计结果。可以看到，当出现不良数据时，RCKF 可以保持更好的鲁棒性。同时，在面对不同量测噪音室，RCKF 可以保持良好的滤波性能。与 RCKF 相比，CKF 无论是鲁棒性还是滤波性能都逊色于 RCKF。出现这种现象的原因是 RCKF 可以通过 M 估计理论对量测噪音的统计特性进行实时更新，发现不良数据并去除其对状态估计的影响。

**（5）　状态估计指标对比**

发电机 2 的状态估计指标如表 2 所示。

**表 2　发电机 2 的动态状态估计指标**

**Tab.2　Dynamic state estimation indexes of the generation 2**

| 类型 | 指标 | 参数 | CKF | 抗差 CKF |
|---|---|---|---|---|
| 高斯白噪音 | $\varepsilon_1$ | $\delta$ | 0.0346 | 0.0161 |
| | | $\omega$ | 0.0075 | 0.0013 |
| | $\varepsilon_2$ | $\delta$ | 0.0078 | 0.0056 |
| | | $\omega$ | 0.0016 | 3.397e-04 |
| 高斯有偏噪音 | $\varepsilon_1$ | $\delta$ | 0.0037 | 0.0018 |
| | | $\omega$ | 0.0071 | 0.0013 |
| | $\varepsilon_2$ | $\delta$ | 0.0142 | 0.0066 |
| | | $\omega$ | 0.003 | 3.389e-04 |
| 拉普拉斯噪音 | $\varepsilon_1$ | $\delta$ | 0.0037 | 0.0018 |
| | | $\omega$ | 0.0070 | 0.0014 |
| | $\varepsilon_2$ | $\delta$ | 0.0141 | 0.0066 |
| | | $\omega$ | 0.0030 | 3.389e-04 |
| 柯西噪音 | $\varepsilon_1$ | $\delta$ | 0.0067 | 0.0019 |
| | | $\omega$ | 0.0074 | 0.0020 |
| | $\varepsilon_2$ | $\delta$ | 0.0121 | 0.0066 |
| | | $\omega$ | 0.0028 | 3.389e-04 |

由上述图表可以看出在面对不良数据时，抗差 CKF 比 CKF 有着更好的鲁棒性。(1) 由估计指标 $\varepsilon_1$ 可知，在面对高斯白噪音污染时，抗差 CKF 的性能相比于 CKF 分别提高了 53.4%和 82%；在面对其它噪音污染时，抗差 CKF 滤波精度仍然优于 CKF。（2）由评估指标 $\varepsilon_2$ 可知，CKF 在面对非高斯白噪音污染时，滤波精度有较大幅度的下滑，而抗差 CKF 对非高斯白噪音污染有着很强的抵抗能力，仍能保持很好的滤波性能和收敛能力。

具体而言，在高斯白噪音仿真中，抗差 CKF 的功角和角速度的滤波精度分别比 CKF 提高了 53.4%和 82%；在高斯有偏噪音仿真中，抗差 CKF 的功角和角速度的滤波精度分别比 CKF 提高了 51.3%和 81%；在拉普拉斯噪音仿真中，抗差 CKF 的功角和角速度的滤波精度分别比 CKF 提高了 51.3%和 80%；在柯西噪音仿真中，抗差 CKF 的功角和角速度的滤波精度分别比 CKF 提高了 71.6%和 72.9%。在所有仿真中不良数据干扰时，抗差 CKF 都具有比 CKF 更好的鲁棒性。

### 3.5 新英格兰 16 机 68 节点系统

#### 3.5.1 算例介绍

该系统由 16 台同步发电机，68 条母线，86 条传输线路组成。在 $t$=1s 时节点 6 发生三相金属性短路故障，1.05s 切除节点 1 的近端故障，1.1s 切除节点 54 的远端故障，仿真持续时间为 10s。

#### 3.5.2 仿真结果

以发电机 1 为例，不同噪声环境的仿真结果如下。

**（1）高斯白噪音**

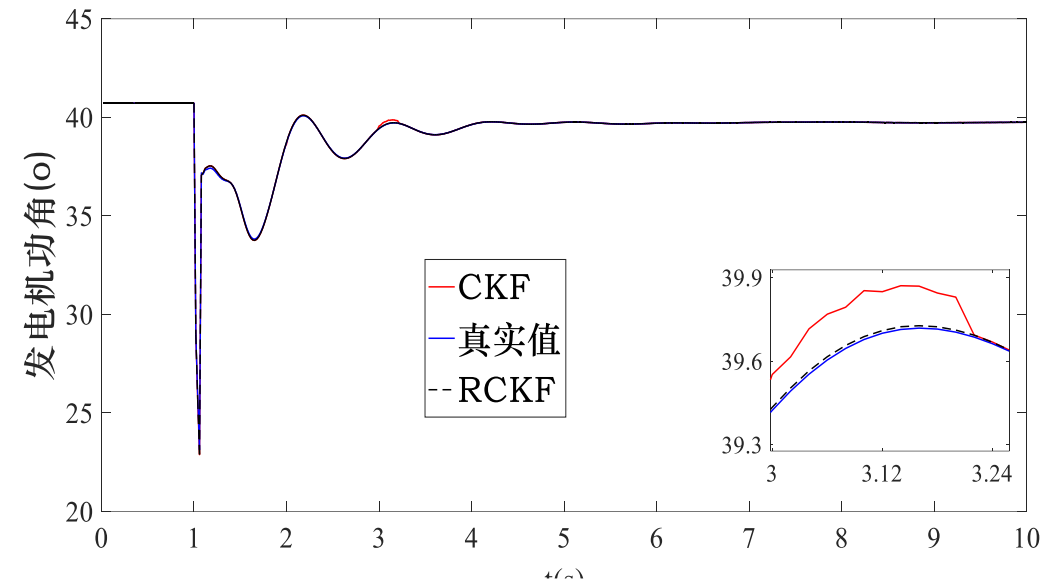

图 9　发电机功角

Fig.9　Estimation of generation power angle

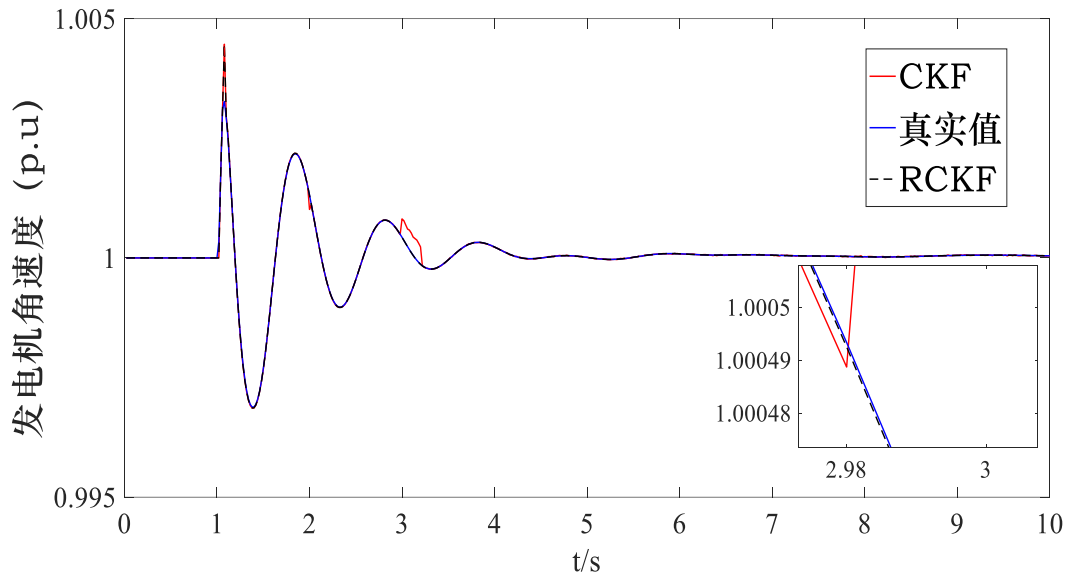

图 10　发电机角速度

Fig.10　Estimation of generation angular velocity

**（2）高斯有偏噪音**

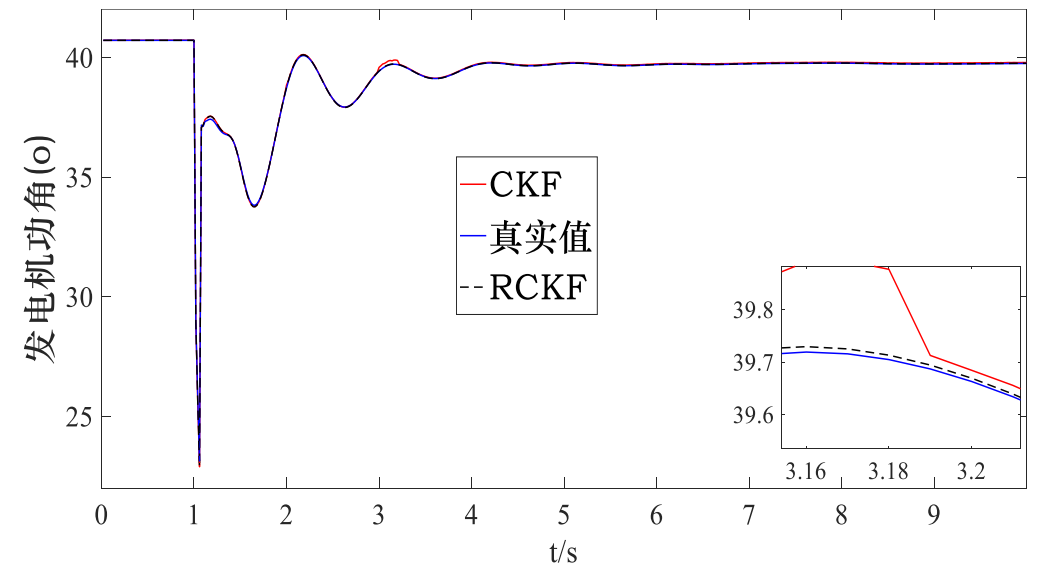

图 11　发电机功角

Fig.11　Estimation of generation power angle

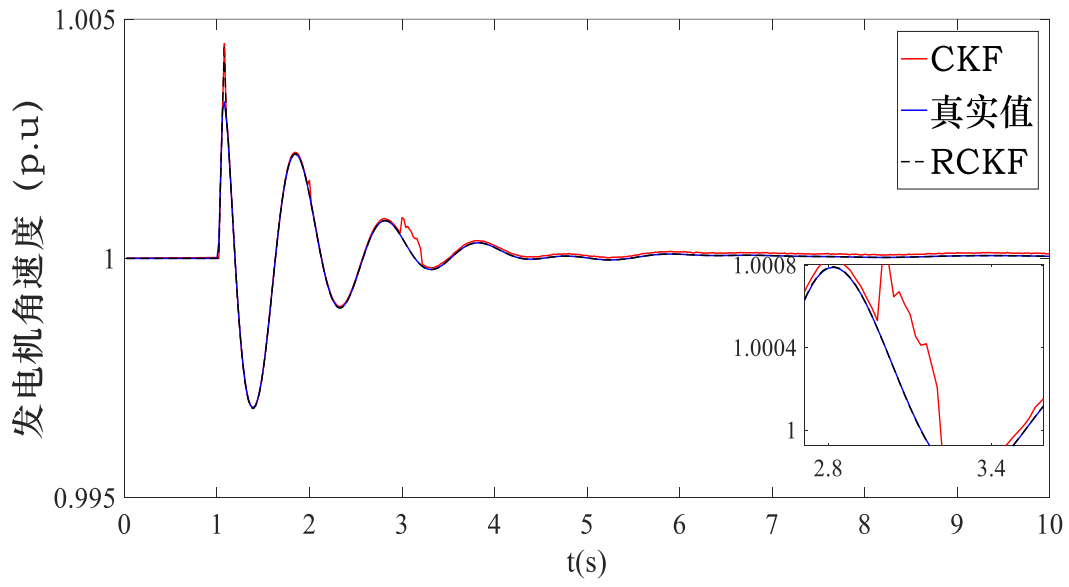

图 12　发电机角速度

Fig.12　Estimation of generation angular velocity

**（3）拉普拉斯噪音**

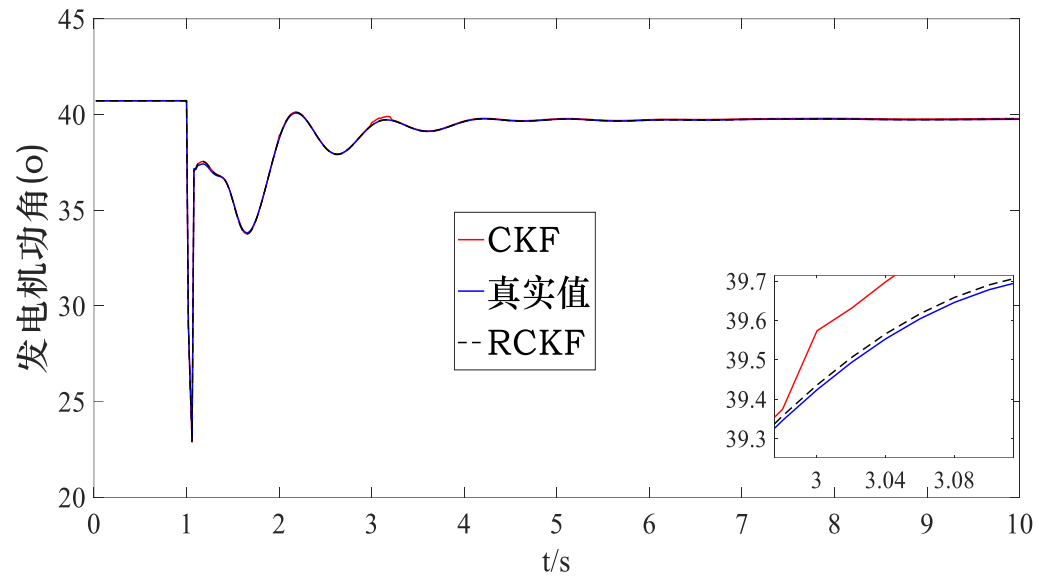

图 13　发电机功角

Fig.13　Estimation of generation power angle

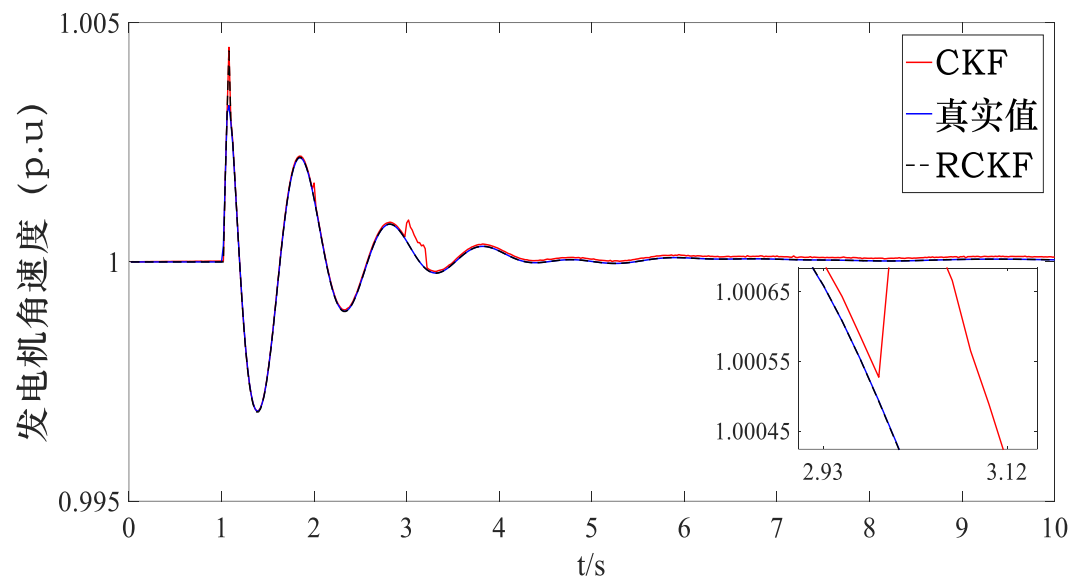

图 14　发电机角速度

Fig.14　Estimation of generation angular velocity

**（4）柯西噪音**

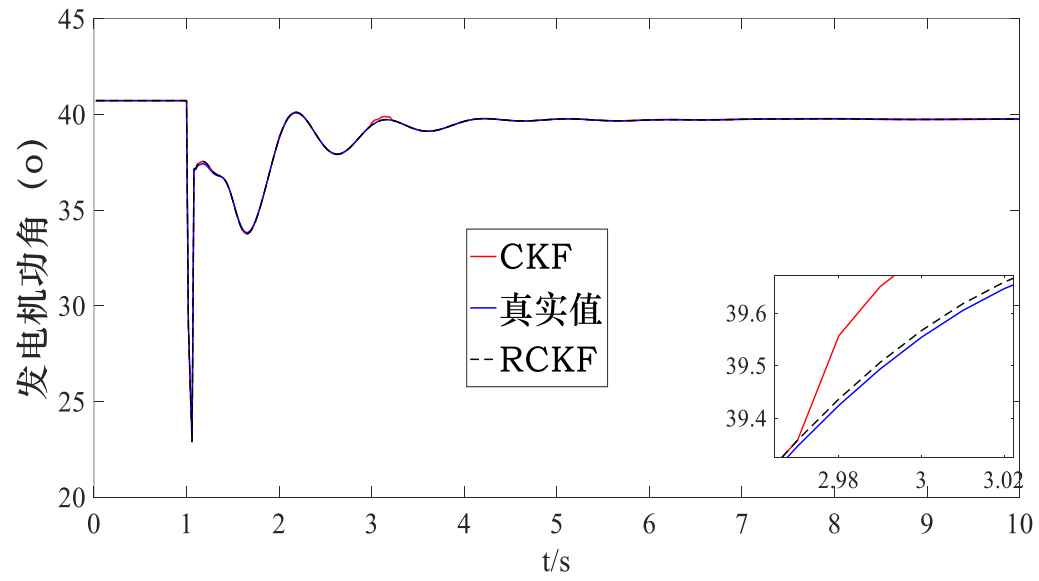

图 15 发电机功角

Fig.15　Estimation of generation power angle

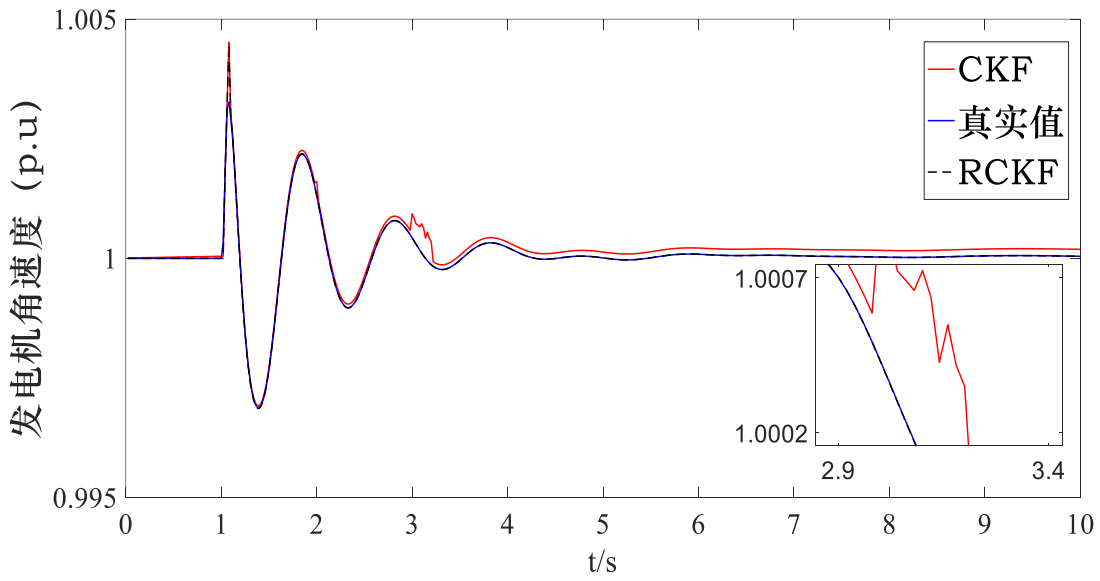

图 16　发电机角速度

Fig.16　Estimation of generation angular velocity

**（5）状态估计指标对比**

发电机 1 的状态估计指标如表 3 所示。

**表 3　发电机 1 的动态状态估计指标**

**Table 3　Dynamic state estimation indexes of generation 1**

| 噪音类型 | 指标 | 参数 | CKF | 抗差 CKF |
|---|---|---|---|---|
| 高斯白噪音 | $\varepsilon_1$ | $\delta$ | 0.0256 | 0.0094 |
| | | $\omega$ | 0.0062 | 0.0018 |
| | $\varepsilon_2$ | $\delta$ | 0.0086 | 0.0031 |
| | | $\omega$ | 0.0012 | 8.397e-05 |
| 高斯有偏噪音 | $\varepsilon_1$ | $\delta$ | 0.0029 | 0.0010 |
| | | $\omega$ | 0.0061 | 0.0017 |
| | $\varepsilon_2$ | $\delta$ | 0.0225 | 0.0064 |
| | | $\omega$ | 0.0024 | 8.177e-05 |
| 拉普拉斯噪音 | $\varepsilon_1$ | $\delta$ | 0.0028 | 0.0010 |
| | | $\omega$ | 0.0061 | 0.0018 |
| | $\varepsilon_2$ | $\delta$ | 0.0228 | 0.0064 |
| | | $\omega$ | 0.0025 | 8.177e-05 |
| 柯西噪音 | $\varepsilon_1$ | $\delta$ | 0.0018 | 9.590e-04 |
| | | $\omega$ | 0.0065 | 0.0031 |
| | $\varepsilon_2$ | $\delta$ | 0.0224 | 0.0064 |
| | | $\omega$ | 0.0015 | 8.173e-05 |

由图 9 至图 16 可知，对于新英格兰 68 节点系统，在面对不良数据干扰和不同的量测噪音时，所提出 RCKF 的鲁棒性和滤波能力均明显优于传统 CKF 算法。

由表 3 可知，对于指标$\varepsilon_1$，在高斯白噪音条件下，RCKF 对发电机功角和角速度的滤波性能分别提高了 63.2%和 70.9%；在高斯噪音条件下，该指标相对提高了 65.5%和 72.1%；在拉普拉斯噪音条件下，该指标相对提高了 64.3%和 70.5%；在柯西噪音条件下，该指标相对提高了 46.7%和 52.3%。

通过观察 CKF 和 RCKF 的性能指标$\varepsilon_2$变化可知，CKF 在面对非高斯白噪音时，滤波性能和估计精度有很大幅度的下滑。而 RCKF 能相对较好地保持其滤波性能和估计精度。综上验证了所提 RCKF 对较大规模系统的适用性。

## 3.6 计算效率比较

上述测试中，两种算法的计算时间对比见表 4.

**表 4　两种算法的计算时间**

**Tab.4　Calculating times of the two algorithms**

| 测试算例 | 噪音 | CKF (ms) | RCKF (ms) |
|---|---|---|---|
| IEEE 9 节点系统 | 高斯白噪音 | 5.35 | 8.16 |
| | 高斯有偏噪音 | 5.38 | 8.18 |
| | 拉普拉斯噪音 | 5.45 | 8.25 |
| | 柯西噪音 | 5.51 | 8.36 |
| 新英格兰 68 节点系统 | 高斯白噪音 | 5.55 | 8.86 |
| | 高斯有偏噪音 | 5.60 | 8.97 |
| | 拉普拉斯噪音 | 5.71 | 9.02 |

| | | |
|---|---|---|
| 柯西噪音 | 5.75 | 9.10 |

由表 4 可知，所提 RCKF 算法的计算时间略高于传统 CKF 算法，但仍然远低于 PMU 采样间隔(20ms). 这表明所提方法足够高效，可以实时追踪发电机的动态状态。

## 4 总结与展望

针对传统CKF缺乏对量测噪音特性的在线自适应能力的问题，本文提出一种基于抗差 CKF 的发电机动态状态估计方法。在两个测试系统的仿真结果表明：在面对同一种噪音污染时，抗差 CKF 的滤波性能均优于 CKF；在面对非高斯白噪音的污染时，CKF 的状态估计效果较差且在个别情况下可能会发散，而抗差 CKF 仍能保持良好的估计精度和收敛效果。

下一步将研究计及模型不确定性、未知输入、信息攻击等场景下的发电机动态状态估计[20]，进而研究基于抗差 CKF 的全系统状态估计。

### 参考文献

[1] 秦晓辉, 毕天姝, 杨奇逊. 基于 WAMS 的电力系统机电暂态过程动态状态估计[J]. 中国电机工程学报, 2008, 28(7): 19-25.
Qin XiaoHui, Bi Tian-Shu, Yang QiXun. Dynamic state estimator based on WAMS during power system transient process[J]. Proceedings of the CSEE, 2008.

[2] Aminifar F, Shahidehpour M, Fotuhi-Firuzabad M, et al. Power system dynamic state estimation with synchronized phasor measurements[J]. IEEE Transactions on Instrumentation & Measurement, 2014, 63(2):352-363.

[3] 罗深增, 李银红, 石东源. 广域测量系统可观性概率评估及其在 PMU 优化配置中的应用[J]. 电工技术学报, 2018, 33(8):1844-1853.
Luo ShenZeng, Li Yinhong, Shi DongYuan. Wide area monitoring system observability probabilistic evaluation and it’s application in optimal PMU placement[J]. Transactions of China Electrotechnical Society, 2018, 33(8):1844-1853.

[4] 肖湘宁, 廖坤玉, 唐松浩,等. 配电网电力电子化的发展和超高次谐波新问题[J]. 电工技术学报, 2018, 33(4):707-720.
Xiao Xiangning, Liao Kunyu, Tang Haosong, et al. Development of power-electronized distribution girds and the new supraharmonics issues[J]. Transactions of China Electrotechnical Society, 2018, 33(4):707-720.

[5] Zhou N, Meng D, Lu S. Estimation of the dynamic states of synchronous machines using an extended particle filter[J]. IEEE Transactions on Power Systems, 2013, 28(4): 4152-4161.

[6] Ghahremani E, Kamwa I. Dynamic state estimation in power system by applying the extended Kalman filter with unknown inputs to phasor measurements[J]. IEEE Transactions on Power Systems, 2011, 26(4): 2556-2566.

[7] 殷实, 谭国俊. 一种基于扩展卡尔曼滤波算法的 MMC 系统故障诊断策略[J]. 电工技术学报, 2016, 31(19):74-84.
Yin Shi, Tan GuoJun. A novel fault diagnosis strategy of MMC system based on EKFA[J]. Transactions of China Electrotechnical Society, 2016, 31(19):74-84.

[8] Zhao J B, Netto M, Mili L. A Robust iterated extended kalman filter for power system dynamic state estimation[J]. IEEE Transactions on Power Systems, 2017, 32(4):3205-3216.

[9] Gultekin S, Paisley J. Nonlinear Kalman filtering with divergence minimization[J]. IEEE Transactions on Signal Processing, 2017, 65(23): 6319-6331.

[10] Qi J, Sun K, Wang J, et al. Dynamic state estimation for multi-machine power system by unscented Kalman filter with enhanced numerical stability[J]. IEEE Transactions on Smart Grid, 2018, 9(2): 1184-1196.

[11] 毕天姝, 陈亮, 薛安成,等. 考虑调速器的发电机动态状态估计方法[J]. 电网技术, 2013, 37(12):3433-3438.
Bi Tianshu, Chen Liang, Xue Ancheng, et al. A dynamic state estimation method considering speed governors[J]. Power System Technology, 2013, 37(12):3433-3438

[12] 孙国强, 王晗雯, 卫志农, 等. 基于无迹粒子滤波算法的发电机动态状态估计[J]. 电力系统自动化，2017, 41(14): 133-139.
Sun Guoqiang, Wang Hanwen, Wei Zhinong, et al. Dynamic state estimation for generators based on unscented particle filtering algorithm[J]. Automation of Electric Power Systems, 2017, 41(14):133-139.

[13] Afshari H H, Gadsden S A, Habibi S. Gaussian filters for parameter and state estimation:A general review of theory and recent trends[J]. Signal Processing, 2017, 135:218-238.

[14] 安军, 杨振瑞, 周毅博, 等. 基于平方根容积卡尔曼滤波的发电机动态状态估计[J]. 电工技术学报，2017, 32(12): 234-240.
An Jun, Yang Zhenrui, Zhou Yibo, et al. Dynamic state estimation for synchronous-machines based on square root cubature Kalman filter[J]. Transactions of China Electrotechnical Society, 2017,32(12): 234-240.

[15] 陈亮, 毕天姝, 李劲松,等. 基于容积卡尔曼滤波的发电机动态状态估计[J]. 中国电机工程学报，2014, 34(16):2706-2713.

Chen Liang, Bi Tianshu, Li Jinsong, et al. Dynamic state estimator for synchronous machines based on cubature Kalman filter[J]. Proceedings of CSEE, 2014, 34(16): 2706-2713.

[16] 毕天姝, 陈亮, 薛安成,等. 基于鲁棒容积卡尔曼滤波器的发电机动态状态估计[J]. 电工技术学报，2016, 31(4):163-169.

Bi Tianshu, Chen Liang, Xue Anchen, et al. Dynamic state estimator for synchronous-machines based on robust cubature Kalman Filter[J]. Transactions of China Electrotechnical Society, 2016,31(4):163-169.

[17] A. Sharma, S.C. Srivastava, and S. Chakrabarti. A Cubature kalman filter based power system dynamic state estimator[J]. IEEE Transaction Instrument Measurement, 2017, 66(8): 2036-2045.

[18] 马安安, 江全元, 熊鸿韬, 等. 考虑量测坏数据的发电机动态状态估计方法[J]. 电力系统自动化, 2017, 41(14): 140-146.

Ma Anan, Jiang QuanYuan, Xiong HongTao, et al. Dynamic state estimation method for generator considering measurement of bad data[J]. Automation of Electric Power Systems, 2017, 41(14):140-146.

[19] Zhou Ning, Meng Da, Huang Zhenyu, et al. Dynamic state estimation of a synchronous machine using PMU data: A comparative study[J]. IEEE Transactions on Smart Grid, 2015, 6(1): 450-460.

[20] 赵欣, 王仕成, 廖守亿,等. 基于抗差自适应容积卡尔曼滤波的超紧耦合跟踪方法[J]. 自动化学报，2014, 40(11):2530-2540.

Zhao Xin, Wang Shicheng, Liao Shouyi, et al. An ultra-tightly coupled tracking method based on robust adaptive cubature kalman filter[J]. Acta Automatica Sinica, 2014,40(11):2530-2540.

---

**作者简介**

李 扬 男，1980 年生，博士，副教授，研究方向为电力系统运行分析与控制。E-mail：liyang@neepu.edu.cn （通信作者）

李 京 男，1990 年生，硕士研究生，研究方向为电力系统运行分析与控制。E-mail：1771104937 @qq. com

陈 亮，男，1984 年生，博士、高工，研究方向为电力系统状态估计。

李国庆，男，1963 年生，博士、教授，研究方向为电力系统运行分析与控制、柔性直流输电技术。